\pdfoutput=1
\documentclass[aps,prl,reprint,groupedaddress,showemail,showpacs]{revtex4-1}

\usepackage{amsthm}
\usepackage{amsmath}
\usepackage{latexsym}
\usepackage{amsfonts}
\usepackage{amssymb}
\usepackage{color}
\usepackage{bbm,dsfont}
\usepackage{graphicx}
\usepackage{enumerate}
\usepackage{hyperref}
\usepackage{subfigure} 
\usepackage{tikz}
\usepackage{placeins}
\usepackage{import}
\usepackage{enumitem}
\usepackage[utf8]{inputenc}
\newcommand\opone{\leavevmode\hbox{\small1\kern-3.8pt\normalsize1}}

\DeclareUnicodeCharacter{00A0}{ }

\usepackage{bbold}
\usepackage[mathscr]{eucal}
\usepackage{ulem}

\newcommand{\W}{\mathscr{W}}
\newcommand{\U}{\mathscr{U}}

\newcommand{\Tr}{\operatorname{Tr}}

\newtheorem{proposition?}{Proposition?}

\theoremstyle{definition}

\newcommand{\ket}[1]{|#1\rangle} 
\newcommand{\bra}[1]{\langle#1|} 
\newcommand{\kb}[2]{|#1\rangle\langle#2|} 

\newcommand{\id}{\mathds{1}} 

\newcommand{\Sys}{\mathcal{S}} 
\newcommand{\Env}{\mathcal{E}} 

\begin{document}


\title{Steering heat engines: a truly quantum Maxwell demon}

\author{Konstantin Beyer} \email{konstantin.beyer@tu-dresden.de}
\affiliation{Institut f{\"u}r Theoretische Physik, Technische
  Universit{\"a}t Dresden, D-01062, Dresden, Germany}

\author{Kimmo Luoma} \email{kimmo.luoma@tu-dresden.de}
\affiliation{Institut f{\"u}r Theoretische Physik, Technische
  Universit{\"a}t Dresden, D-01062, Dresden, Germany}

\author{Walter
  T. Strunz} \email{walter.strunz@tu-dresden.de} \affiliation{Institut
  f{\"u}r Theoretische Physik, Technische Universit{\"a}t Dresden,
  D-01062, Dresden, Germany}

\date{\today}

\begin{abstract}
We address the question of verifying the quantumness of thermal
machines. A Szilárd engine is truly quantum if its work output cannot
be described by  {a local hidden state (LHS) model, i.\,e. an objective local statistical ensemble.}
Quantumness in this scenario is revealed by a steering-type inequality
which bounds the classically extractable work.
A quantum Maxwell demon can violate that inequality by exploiting
quantum correlations between the work medium and the thermal
environment. While for a classical Szilárd engine
an objective description of the  medium always exists, any such description can be
ruled out by a steering task in a truly quantum case.
\end{abstract}


\maketitle


\paragraph*{Introduction}
--- Experimental progress has led to unprecedented possibilities of
preparation, control, and measurement of small quantum systems, where
quantum and thermal fluctuations have to be considered on equal
footing. In particular, fundamental concepts of thermodynamics have
been revisited from a quantum point of view. This has led to a quantum
interpretation of thermal
states~\cite{SrednickiChaosquantumthermalization1994,DAlessioquantumchaoseigenstate2016,GogolinEquilibrationthermalisationemergence2016,Gooldrolequantuminformation2016},
the development of quantum fluctuation theorems~\cite{AbergFullyQuantumFluctuation2018,Elouardrolequantummeasurement2017,AlhambraFluctuatingWorkQuantum2016,ManzanoNonequilibriumpotentialfluctuation2015,
  LeggioEntropyproductioninformation2013,HekkingQuantumJumpApproach2013,ChetriteQuantumFluctuationRelations2012,HorowitzQuantumtrajectoryapproachstochastic2012,
  CampisiColloquiumQuantumfluctuation2011,EspositoNonequilibriumfluctuationsfluctuation2009,TalknerFluctuationtheoremsWork2007} and the concepts of quantum heat
engines~\cite{Gevaquantummechanicalheat1992,FeldmannQuantumfourstrokeheat2003,HeThermalentangledquantum2012,
  ParkHeatEngineDriven2013,ZhangQuantumOttoheat2014,MohammadyquantumSzilardengine2017,ZhaoEntangledquantumOtto2017,
  ThomasThermodynamicsnonMarkovianreservoirs2018,PezzuttooutofequilibriumnonMarkovianquantum2019}.
One of the key points in these investigations is the question what is
fundamentally quantum about these extensions. For instance, whether
and how is  a quantum heat engine  qualitatively and quantitatively different from
its classical counterpart? Is quantumness useful in thermodynamics?
Influences of quantum features like coherence~\cite{ScullyQuantumheatengine2011,UzdinEquivalenceQuantumHeat2015},
discord~\cite{DillenschneiderEnergeticsquantumcorrelations2009} and
entanglement~\cite{AllahverdyanBreakdownLandauerbound2001,rioThermodynamicMeaningNegative2011} on the efficiency of quantum engines have been
reported, which show that the answer can be positive under suitable
conditions. {However, other investigations show that quantumness can
even be a hindrance for efficient thermal machines, which can be
regarded as classical supremacy in such situations~\cite{KarimiOttorefrigeratorbased2016,BrandnerPeriodicthermodynamicsopen2016,PekolaSupremacyincoherentsudden2018}. }

In this Letter we want to address quantumness of thermal machines
from a different perspective.
We consider a heat engine truly quantum if its work output cannot be
explained by a  {local hidden state (LHS) model, i.\,e. by a local statistical model.} Even though the issue of
hidden classicality is fundamental to quantum information, it only
rarely appears in the context of quantum
thermodynamics~\cite{FriedenbergerWhenquantumheat2017}. In this Letter
we give a verifiable criterion for the quantumness of thermodynamical
systems, indicating the lack of a classical statistical
description. Most remarkably, the classicality sets an upper bound on
the extractable work for certain scenarios.

\paragraph*{Quantum Szilárd engine}
--- The prototypical example we want to study is a quantum modification of
the Szilárd engine~\cite{SzilardueberEntropieverminderungthermodynamischen1929,MaruyamaColloquiumphysicsMaxwell2009}.
The classical version  consists of a single atom in a box which is in
contact with a thermal bath. In equilibrium the atom is in a Gibbs state, a
statistical mixture of different phase space points. For work
extraction, the demon has knowledge about the microstate of the
system.

So far, quantum versions of this heat engine have been investigated using different underlying
systems~\cite{KimQuantumSzilardEngine2011, ParkHeatEngineDriven2013,Faistminimalworkcost2015,MohammadyquantumSzilardengine2017,
  ElouardExtractingWorkQuantum2017}. In
these examples the demon performs quantum measurements on the work
medium, acquiring information about local properties of the heat
engine only.
Here, we want to
exploit the fact that such a local thermal state may arise naturally from
a global entangled state of the work medium and its environment, as
supported, for instance, by the eigenstate thermalization
hypothesis~\cite{SrednickiChaosquantumthermalization1994,DAlessioquantumchaoseigenstate2016,GogolinEquilibrationthermalisationemergence2016,Gooldrolequantuminformation2016,FaistFundamentalWorkCost2018}.
In contrast to previous proposals, the demon obtains her
information from measurements on the environment rather than the work
medium~\cite{MorrisAssistedWorkDistillation2019,ManzanoOptimalWorkExtraction2018}.
A truly quantum Szilárd engine can be revealed by deriving local work
extraction bounds which cannot be violated by any  {local statistical ensemble description, that is a LHS model.} 
These bounds do neither rely on the knowledge
about  the shared system-environment state, nor on any assumptions about the properties
of the environment (semi-device-independent).

\paragraph*{Work medium}
---Let us assume that the work medium is a finite
quantum system $\Sys$ with Hamiltonian $H_\Sys$. 
Its Gibbs state reads
\begin{align}
  \label{eq:3}
  \rho_\Sys^{\textrm{Gibbs}} = \sum_{i} \frac{e^{-\beta E_i}}{Z}
  \kb{i}{i} = \sum_{i} p_i \kb{i}{i} ,
\end{align}
where $\beta = \frac{1}{k_B T}$, $E_i$ is the energy of the $i$th
energy eigenstate $\ket{i}$ and $Z = \sum_{i} e^{-\beta
  E_i}$. Locally, the Gibbs state can be seen as a statistical mixture of the
energy eigenstates but it can equally be decomposed into infinitely
many other ensembles $\mathcal{D} = \{ p_k; \rho_k  \}$ with
$\sum_k p_k \rho_k = \rho_{\Sys}^{\textrm{Gibbs}}$, $p_k\geq 0$ and
$\sum_k p_k =1$. Any such
decomposition can be given by an extension to a bipartite state
$\rho_{\Sys\Env}$ of system and environment, with
$\rho_{\Sys}^{\textrm{Gibbs}}=\Tr_{\Env}\{\rho_{\Sys\Env}\}$, and
a local POVM $\{ M_k \}$ on $\Env$, such that $p_k = \Tr\{(\id \otimes
M_k)\rho_{\Sys\Env}\}$ and $\rho_k =\Tr_\Env\{(\id \otimes M_k)\rho_{\Sys\Env}\}$.

To investigate the difference of a classical and a quantum
Szilárd engine we introduce Alice and Bob. Alice is a demon who can
prepare many copies of the global state $\rho_{\Sys\Env}$. While she can perform measurements on $\Env$, she does not act directly on $\Sys$ since this would in
general disturb the local thermodynamical situation~\footnote{For example all measurements, apart from
those which are diagonal in the energy basis, would change the average
inner energy of the system. Such a measurement fueled engine
 has been described in~\cite{ElouardExtractingWorkQuantum2017}. There
 the quantum measurements inject energy which can be extracted as work
afterwards.}.
Bob has access only  to
the system $\Sys$ and would like to extract work from its Gibbs
state. As shown in~\cite{MorrisAssistedWorkDistillation2019} any
non-product joint state $\rho_{\Sys\Env} \neq \rho_\Sys \otimes
\rho_\Env$ allows Bob to extract work from $\Sys$ if Alice performs
suitable measurements on $\Env$ and communicates classically with him.

\paragraph*{Work extraction scenarios}
Bob wants to extract work from a certain decomposition $\mathcal{D} =
\{p_k; \rho_k \}$ of his local
Gibbs state. We can describe this work extraction as a transfer of
energy due to a  {suitable} coupling between the work medium  $\Sys$ and a work storage
system $\W$  {with Hamiltonians $H_\Sys$ and $H_\W$, respectively}~\cite{AlhambraFluctuatingWorkQuantum2016,AbergFullyQuantumFluctuation2018,BrandaoResourceTheoryQuantum2013,abergCatalyticCoherence2014}. 
The total Hamiltonian is given by $H = H_\Sys \otimes \id +
\id \otimes H_\W$. Following the reasoning of~\cite{AlhambraFluctuatingWorkQuantum2016}
the coupling of $\Sys$ and $\W$ has to be a unitary transformation
that conserves the total energy {independently of the initial state of the work storage $\rho_\W$}. 
 {Further constraints ensuring that only work and no heat is transfered are presented in the 
Supplemental Material~\footnote{See Supplemental Material, which includes Refs.~\cite{masanesGeneralDerivationQuantification2017,lorenzoCompositeQuantumCollision2017,filippovDivisibilityQuantumDynamical2017,campbellSystemenvironmentCorrelationsMarkovian2018,cavalcantiExperimentalCriteriaSteering2009,ciccarelloCollisionModelsQuantum2017}.}
For each state $\rho_k$ in $\mathcal{D}$ Bob can find a
suitable unitary $\U_k$ which transfers a non-negative amount of energy from $\Sys$ to $\W$. The
average work extraction associated with this process is given by 
\begin{align}
  \label{eq:5}
  \Delta W_k = \Tr\{\id\otimes H_\W(\U_k\, \rho_k\otimes\rho_\W \, \U_k^\dagger
  - \rho_k \otimes\rho_\W)\},
\end{align}
where $\rho_\W$ is the initial state of the work storage.  {Due to energy conservation, the change of the inner energy in $\Sys$ is 
$\Delta E_k = -\Delta W_k$} and $\Delta
W_k \leq \Tr\{H_\Sys \rho_k  \}$. The average work Bob
can extract from $\mathcal{D}$ by using the set of unitaries $U = \{\U_k\}$ is then given by $\overline{W} = \sum_k
p_k \Delta W_k$. 

Special cases are pure state decompositions $\mathcal{D}^{\textrm{pur}} = \{p_k; \kb{\phi_k}{\phi_k}\}$.
If Bob knows the pure state $\ket{\phi_k}$ of his system, {the largest possible amount of work
can only be extracted if he performs a} suitable \textit{local} unitary operation $\U_k^\Sys$ on $\Sys$ which acts as
$\U_k^\Sys \ket{\phi_k} = \ket{0}$, where $\ket{0}$ is the ground state of
the local Hamiltonian $H_\Sys$ whose energy we set to $E_0=0$~\cite{perarnau-llobetExtractableWorkCorrelations2015}. As shown
in~\cite{AlhambraFluctuatingWorkQuantum2016}, such a local unitary {$\U_k^\Sys$} can indeed
always be implemented by an energy conserving global coupling {$\U_k$ between $\Sys$ and} $\W$
but requires the work storage to be initialized in a pure state $\rho_\W$ which is a coherent superposition of energy eigenstates states of $H_\W$.
Thus, energy measurements on $\W$ will in general yield probabilistic
outcomes~\cite{tajimaUncertaintyRelationsImplementation2018}. {However, Bob is only interested in the average work output under the given $\U_k$, which
is, due to the energy conservation, given by the negative energy change in $\Sys$, that is}    
$\Delta W_k = {-\Delta E_k =} \bra{\phi_k}H_\Sys\ket{\phi_k} -
\bra{0}H_\Sys\ket{0} =\bra{\phi_k}H_\Sys\ket{\phi_k} $.
Accordingly, if Alice can provide the pure state decomposition $\mathcal{D}^{\textrm{pur}}$, Bob can
extract on average $\overline{W}^{\textrm{pur}} =  \sum_k p_k
\bra{\phi_k}H_\Sys\ket{\phi_k} = \langle H_\Sys
\rangle_{\rho_{\Sys}^{\textrm{Gibbs}}}$ which is, not surprisingly,
the inner energy of the work medium. Thus, equivalently to the
classical case, full knowledge about the state of the system allows
for maximal {work} extraction. If Alice cannot announce correct pure
states to Bob, the output will be $\overline{W} < \overline{W}^{\textrm{pur}}$. 
In fundamental contrast to the classical Szilárd scenario, a Gibbs
state of a quantum system allows for infinitely many different ensembles
of pure states.

Truly quantum features can be revealed when Bob wants to extract work from
different decompositions $\mathcal{D}_n$. For each
decomposition he has a suitable set of unitaries $U_n =  \{ \U^n_{k_n}\}$ as
described above. He chooses randomly with probabilities $c_n$ one of the sets and asks Alice
which unitary out of the particular set $U_n$ he should perform to extract
the maximal amount of work. Depending on how well Alice can produce
the desired decompositions, Bob will extract on average $\overline{W}
\leq \sum_n c_n \overline{W}_n$.

The question now arises, under which conditions Bob can be sure that
his Szilárd engine is truly quantum. He has no access
to the global state $\rho_{\Sys\Env}$ and, therefore, cannot check
whether the state is quantum correlated. The only {information} he gets from Alice {is} which unitary $\U^n_{k_n}$ he should use if he asks her for
the decomposition $\mathcal{D}_n$. Accordingly,
Bob has to certify quantumness without any assumptions about
the properties (for example the Hilbert space) of the environment
$\Env$. Such a semi-device-independent verification task is
called quantum steering~\cite{WisemanSteeringEntanglementNonlocality2007,JonesEntanglementEinsteinPodolskyRosencorrelations2007}. Successful steering has important implications on the
objectivity~\cite{wisemanAreDynamicalQuantum2012} of the local state in the system.
In a classical scenario the system state is always objective, though
unknown to Bob as long as the demon does not share her knowledge with him. In the quantum case, in general, it makes no sense to assign objective system states at all, as long as
no observation of the environment is made. Particularly, a thermalized
quantum system is \textit{not} in one of its energy
eigenstates and does \textit{not} fluctuate between them while time is evolving, if
these fluctuations are not given relative to measured states of the
bath~\footnote{Of course, objective states can be assigned by direct measurements on the system, but this case is not
  considered in our setting. Furthermore, we should note that ensemble
  representations of mixed states can be computationally useful in
  order to calculate averaged quantities more efficiently.}.
For a closer look on how steering can rule out objective quantum
dynamics see~\cite{wisemanAreDynamicalQuantum2012,DaryanooshQuantumjumpsare2014,BeyerCollisionmodelapproachsteering2018}. In
our Szilárd scenario we can use these ideas as follows:
If a local {objective} statistical description of Bob's system $\Sys$ holds,
it can be represented by {a local hidden state (LHS) model $\mathcal{F} = \{p_\xi; \rho_\xi\}$~\cite{JonesEntanglementEinsteinPodolskyRosencorrelations2007}. 
The hidden states $\rho_\xi$ are distributed randomly according to their probabilities $p_\xi$. 
Locally the Gibbs state in Bob's system has to be recovered
\begin{align}
  \label{eq:11}
  \rho^{\textrm{Gibbs}}_\Sys = \sum_\xi p_\xi \rho_\xi.
\end{align}}
Thus, among all the copies of his local state, a fraction $p_{{\xi}}$ will be in state $\rho_{{\xi}}$. Bob does not know
which state he has for a particular copy but he {can assume that, if the LHS model holds, the best knowledge Alice can
 possibly have about his system is the particular hidden state for each of his copies. Therefore, any decomposition Alice can provide 
 has to be either the LHS ensemble $\mathcal{F}$ itself or a coarse graining of the latter~\cite{JonesEntanglementEinsteinPodolskyRosencorrelations2007}. 
 However, this does not mean that the real state $\rho_{\Sys\Env}$ shared by Alice and Bob has to be separable. It only means that
 Bob could explain his statistics also by a state without quantum correlations.}
A truly quantum Szilárd engine can therefore be defined by the condition
$\overline{W} > \overline{W}_{\textrm{cl}}$, that is, Bob's average
work output is larger  than what could be obtained
from a {state which can be described by a LHS ensemble $\mathcal{F}$}. Clearly, the
work output of a single decomposition {$\mathcal{D}$} can always be explained by a
classically {correlated} state {because we can always identify $\mathcal{D} = \mathcal{F}$.} Bob needs at least two
different sets of unitaries $U_n$.

We should note that the observables on Bob's side needed to perform a
steering task are represented by the work extraction. In order to
determine the average energy transferred to the work storage he has to
measure $\W$ {in its energy basis}. {According to Naimark's dilation theorem,} this
measurement, together with a
unitary $\U_k$, defines a POVM on $\Sys$. The set of POVMs which can
be implemented by the described work extraction scenario is strictly
smaller than the set of all local POVMs on $\Sys$. For
example, the only implementable projective measurement is the one
diagonal in the energy eigenbasis of $H_\Sys$. It is an open question
whether the work extraction POVMs can demonstrate steering for any
steerable state $\rho_{\Sys\Env}$ which respects the local Gibbs state.

Whether the work output on Bob's side can also be provided by a
classical demon is in general not trivial to answer. As in a standard
steering scenario a suitable inequality has to be derived which
depends on the properties of the work medium $\Sys$ and the work
extracting unitaries $\{\U_k^n\}$. It is crucial for quantum steering
that the inequality does not depend on the part $\Env$ which is
inaccessible for Bob.

\paragraph*{Qubit work medium}
--- To illustrate the concept we consider a qubit work medium $\Sys$
with local Hamiltonian $H_\Sys = \kb{1}{1}$. Its thermalized Gibbs
state is given by
\begin{align}
  \label{eq:1}
  \rho_{\Sys}^{\textrm{Gibbs}} = \frac{1+\eta}{2} \kb{1}{1} +\frac{1-\eta}{2} \kb{0}{0},
\end{align}
with $\eta = \frac{e^{-\beta}-1} {e^{-\beta}+1}<0$ and $\beta =
\frac{1}{k_B T}$.
As stated above, Bob needs at least two different sets of work
extracting unitaries to verify a quantum Szilárd engine. Let us assume
that he would like to extract work from two dichotomic pure state
decompositions $\mathcal{D}_1$ and $\mathcal{D}_2$. The first one is a
decomposition into energy eigenstates $\{\ket{0}, \ket{1}\}$, the
second one is given by the two Bloch vectors $\vec{r}_\pm = (\pm \sqrt{1- \eta^2},0,\eta)$.
The local unitaries for
$\mathcal{D}_1$ are $\U_z^1 = \sigma_x$ for the state $\ket{1}$ and
$\U_z^0 = \id$ for the state $\ket{0}$. For $\mathcal{D}_2$ the
suitable unitaries $\U_x^\pm$ are rotations around the $y$-axis about an angle
$\alpha = \pm \arctan(\frac{\eta}{\sqrt{1-\eta^2}})$~(see
Fig.~\ref{fig:work-extraction}).
Accordingly, Bob needs two different kinds of work extraction
devices. We represent them by red and blue cells which both have two
buttons to trigger the different work extraction unitaries and measure
the work storage $\W$ in the energy basis~(Fig. \ref{fig:work-extraction}). In each
cell Bob can place one qubit. The red cells can perform  $\U_z^1$ and $\U_z^0$, the blue cells apply either $\U_x^+$ or
$\U_x^-$. In the Supplemental Material~\cite{Note2} we construct an explicit model, how the energy conserving
unitaries can be realized by using only two qubit interactions.

Let us first assume that Alice prepares
the global state $\rho_{\Sys\Env}^{\mathcal{D}_1} =  \frac{1+\eta}{2} \kb{1}{1} \otimes\kb{1}{1}                       
+ \frac{1-\eta}{2}\kb{0}{0}\otimes\kb{0}{0}$, compatible with the
local Gibbs state. Bob places his qubit
into a red cell and asks Alice which button he should press. Alice measures $\Env$ in the $\sigma_z$-basis and tells Bob to
press the button 1 if the outcome is 1 and button 0 if the outcome is
0. The cell will apply either $\U_z^1$ or $\U^0_z$.
On average --- Bob has many red cells which he
wants to charge --- he will extract $\overline{W}_z =
\frac{1+\eta}{2}$ because Alice tells him to press button~1 with
probability $p_z^1 = \frac{1+\eta}{2}$.

If Bob wants to charge his blue cells, Alice could help him by
preparing the state $\rho_{\Sys\Env}^{\mathcal{D}_2} =  \rho_{+} \otimes\kb{1}{1}                       
+ \rho_{-}\otimes\kb{0}{0}$, where $\rho_\pm$ are the density
matrices corresponding to the Bloch vectors $\vec{r}_\pm$. Locally, the
Gibbs state is again recovered. Depending
on her outcome, Alice tells Bob to press either the button which
applies $\U_x^+$ or $\U_x^-$ (see Fig.~\ref{fig:work-extraction}). On average Bob can again extract
$\overline{W}_x = \frac{1+\eta}{2} = \overline{W}_z$. Thus, the
decomposition $\mathcal{D}_1$ into energy eigenstates is by no means
better for the work extraction than decomposition $\mathcal{D}_2$.
Both  $\rho_{\Sys\Env}^{\mathcal{D}_1}$ and
$\rho_{\Sys\Env}^{\mathcal{D}_2}$ are separable and describe situations where Alice
exploits only classical correlations. We call
such a demon a classical one because the same result can be obtained
from a local statistical model for $\rho_\Sys$ without any
reference to a global quantum state $\rho_{\Sys\Env}$.
\begin{figure}
    \includegraphics[width=\linewidth]{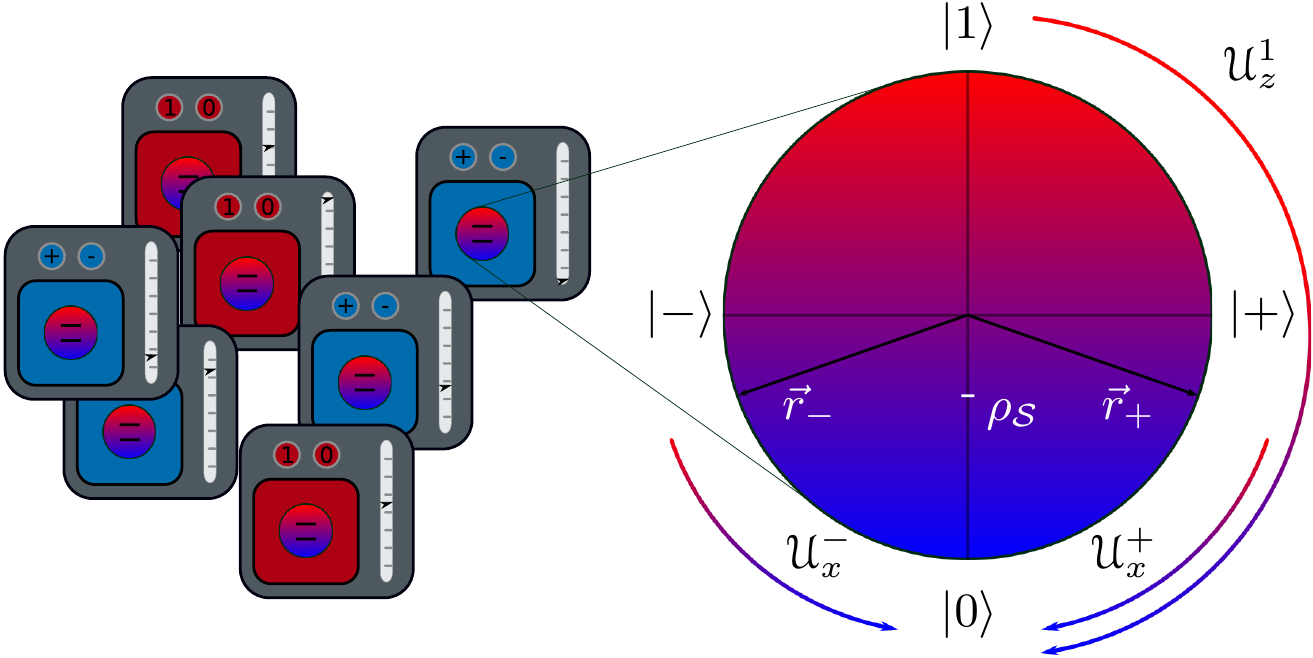}
    \caption{Work extraction. Bob has blue and red work
      extraction cells with locally thermal qubits. The
      reduced state $\rho_\Sys$ can be decomposed into statistical
      mixture of $\ket{1}$ and $\ket{0}$ or
      the two states given by the Bloch vectors
      $\vec{r}_\pm$. Different unitaries can be used to bring the
      states to the ground state extracting some work. The red cells
      can apply the $\U_z^1$ and the identity operation. Qubits in the blue cells can
      be manipulated by two unitaries $\U_x^\pm$. Bob gets the
      information which unitary he should use from Alice. After the
      coupling process the work storage is measured in its energy basis.}\label{fig:work-extraction}
\end{figure}

\paragraph{When is the demon really quantum?}
--- In the remainder we will consider the case where Bob
would like to charge both the red and the blue cells. He has $N$
blue cells, $M$ red cells and $N+M$ thermal qubits. The ratio between red and blue cells is given by $c =
\frac{N}{M}$. First, he distributes the qubits over the cells which
fixes the decomposition he needs to extract maximal work with any
given cell. Subsequently, he announces the
color of each single cell to Alice and asks her which
button he should press. In the end he can read off the extracted work
from each work meter and average over them to see how efficient the
procedure has been. If Alice's knowledge for each single qubit is
described by $\rho_{\Sys\Env}^{\mathcal{D}_1}$ or
$\rho_{\Sys\Env}^{\mathcal{D}_2}$, his average work output in the
limit $N\rightarrow \infty$ can never
reach the optimal $\overline{W}_{\textrm{opt}} = \frac{1+\eta}{2}$ (we
assume that $c > 0$ is kept constant). The blue and the red cells are not compatible
with the same statistical mixture {of states}. On the other hand the
entangled state of the form
\begin{align}
  \label{eq:7}
  \ket{\Psi}_{\Sys\Env} = \sqrt{\frac{1+\eta}{2}} \ket{1}_\Sys \otimes \ket{
  1}_\Env +  \sqrt{\frac{1-\eta}{2}}\ket{0}_\Sys \otimes\ket{0}_\Env.
\end{align}
would do the job. Alice could measure either $\sigma_z$ or $\sigma_x$ depending on the color
of the cell for which Bob would like to know which button he should
press. If Alice is indeed a demon who can prepare Bob's thermal state
to be the partial trace of a pure entangled state, he can
extract the optimal average $\overline{W}_{\textrm{opt}}$.

We will now calculate which average work Bob can maximally obtain {if a LHS model holds.} 
The best Alice can do if she knows the state $\rho_\xi$ for each cell is to tell Bob which
button he should press in order to obtain the maximal work output.
Accordingly, for the red cells Alice would tell him to press the
button triggering  $\U_z^1$ whenever $z_\xi = \Tr\{\sigma_z \rho_\xi\}
>0$ for the state in this cell. The average work output for the red cells will then be 
\mbox{$\overline{W}_z =\frac{1}{2}\sum_\xi p_\xi (|z_\xi| +
  \eta)$}~\cite{Note2}. For the blue cells Alice announces the button for $\U_x^+$ if
  $x_\xi = \Tr\{\sigma_x \rho_\xi\} >0$ and the button for  $\U_x^-$ if
  $ x_\xi \leq 0$.  On average the blue cells will then reach
  \mbox{$\overline{W}_x = \frac{1}{2}(\eta + \eta^2 + \sum_\xi p_\xi
  |x_\xi|\sqrt{1-\eta^2})$}~\cite{Note2}.
The work average over all cells which can be expected for a {LHS} 
model is, thus, given by $\overline{W}_{\textrm{cl}}  =
\frac{\overline{W}_z + c \, \overline{W}_x}{1+c}$.
For the choice $c = \frac{1}{\sqrt{1-\eta^2}}$ we can bound the
average work for any {LHS} model by~\cite{Note2}
\begin{align}
  \label{eq:23}
  \overline{W}_{\textrm{cl}} \leq \frac{\eta  \left(\sqrt{1-\eta ^2}+\eta +1\right)+\sqrt{2-2 \eta ^2}}{2 \left(\sqrt{1-\eta ^2}+1\right)}.
\end{align} 
If Bob extracts an average
work beyond the classical limit  $\overline{W}_{\textrm{cl}}$ he can
be satisfied that Alice is indeed a quantum demon who exploits
non-classical correlations. Any local statistical model has to be
discarded in this case.

If Alice can, for example, indeed prepare the entangled state
~(\ref{eq:7}), the work value which can be reached is $\overline{W}_{\textrm{qu}} =\overline{W}_{\textrm{opt}}= \frac{1 + \eta}{2}$.
Thus, Alice can violate the inequality for $\eta >-\sqrt{2
  (\sqrt{2}-1)} \approx -0.91$, that is, for temperatures $k_B T >
0.33$. For the case of infinite temperature
($\eta \rightarrow 0$) the steering inequality simplifies to
$\overline{W}_{\textrm{cl}} \leq \frac{1}{2\sqrt{2}}$, while
$\overline{W}_{\textrm{opt}}=\frac{1}{2}$.

For an intermediate regime we can consider
the  following mixture
\begin{align}
  \label{eq:29}
  \rho_{\Sys\Env} = q\,\rho_{\textrm{qu}} + (1-q) \rho_{\textrm{cl}},
\end{align}
where $\rho_{\textrm{qu}} = \kb{\Psi}{\Psi}_{\Sys\Env}$ as in
Eq.~(\ref{eq:7}) and $\rho_{\textrm{cl}}$ is the classically
correlated state \mbox{$\rho_{\textrm{cl}} = \frac{1+\eta}{2}
                         \kb{1}{1}_\Sys\otimes\kb{1}{1}_{\Env}  +
                         \frac{1-\eta}{2}\kb{0}{0}_\Sys\otimes\kb{0}{0}_{\Env}$}.                         
The parameter $q$ tunes between the fully quantum case ($q = 1$) and
the scenario which represents a classical Szilárd demon ($q =0$).
The extractable work in the blue cells is now \mbox{$\overline{W}_{x} = \frac{1}{2}(q + \eta + \eta^2-q
\eta^2)$}~\cite{Note2}.
Fig.~\ref{fig:non-cl} shows the relation between the non-classicality of the demon and the two parameters $\eta$ and $q$. For
parameters above the red line Alice can demonstrate that she is a
quantum Szilárd demon.
\begin{figure}
  \includegraphics[width=.65\linewidth]{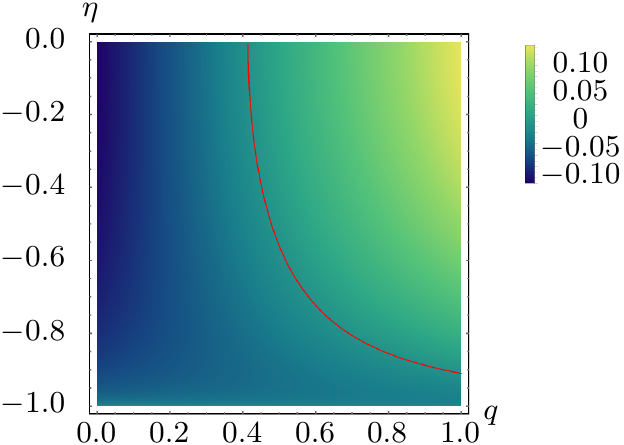}
    \caption{The plot shows by how much Alice can violate the
      classical work bound. For parameters in the region below the red
      line Bob can explain the extracted work with a local {hidden state} 
    model. }\label{fig:non-cl}
  \end{figure}

We should note that the steering inequality~(\ref{eq:23}) is not ideal
for the detection of non-classical correlations in the state
$\ket{\Psi}_{\Sys\Env}$. {It is well known that any pure entangled state is steerable~\cite{WisemanSteeringEntanglementNonlocality2007},
thus,  ${\ket{\Psi}}_{\Sys\Env}$ is steerable for any $-1<\eta<1$. More generally, any non-zero temperature Gibbs state is mixed and therefore has 
an entangled and steerable purification.} 
However, we have restricted Bob's observables to a
special class of operations, namely the work extraction. It is
therefore not surprising that the given inequality cannot detect
every steerable state. {We could improve the bound by adding additional work
extraction options on Bob's side, but this does not add anything
conceptually new to the framework. Furthermore, motivated by the concept of 
a Szilárd engine, the inequality is based on the assumption
that Bob's reduced state is indeed a Gibbs state. This property can of course be locally verified by Bob.}
It has to be emphasized that the construction of the steering
inequality only depends on the device-dependent part of the steering
task, such as the Hilbert space of Bob's system
$\Sys$ and the work extracting operations he uses. There
are no assumptions made about the structure of the environment or the
operations Alice performs.

\paragraph*{Conclusions}
  In this Letter we have shown how the concept of quantum steering can be applied to quantum thermodynamics
  in order to verify quantumness.
  The violation of a steering inequality is connected to the
  macroscopic average work. The use of a quantum steering task for the verification of
  quantumness is motivated by the asymmetric setting in quantum heat
  engines. The work system under control is taken to be the
  device-dependent part in the scenario, whereas the environment is
  treated device-independently.

  Our concept is of particular interest for the investigation of
  bath-induced fluctuations in quantum thermodynamics. A violation of
  the steering inequality rules out any possible objective (though statistical) description
  of fluctuations in the system. Notably, the assumption that a system
  fluctuates between its energy eigenstates is not valid if genuine
  quantum correlations are taken into account. Statements about the fluctuations in the
  system can only be made with respect to the observed fluctuations of
  the environment which will depend on how the environment is
  measured.

  \begin{acknowledgments}
  The authors would like to thank Dario Egloff for
  helpful comments on the manuscript.
\end{acknowledgments}

\bibliography{bibliography}
\end{document}



\title{Steering heat engines: a truly quantum Maxwell demon --- Supplemental Material}

\author{Konstantin Beyer} \email{konstantin.beyer@tu-dresden.de}
\affiliation{Institut f{\"u}r Theoretische Physik, Technische
  Universit{\"a}t Dresden, D-01062, Dresden, Germany}

\author{Kimmo Luoma} \email{kimmo.luoma@tu-dresden.de}
\affiliation{Institut f{\"u}r Theoretische Physik, Technische
  Universit{\"a}t Dresden, D-01062, Dresden, Germany}

\author{Walter
  T. Strunz} \email{walter.strunz@tu-dresden.de} \affiliation{Institut
  f{\"u}r Theoretische Physik, Technische Universit{\"a}t Dresden,
  D-01062, Dresden, Germany}

\date{\today}

\begin{abstract}
\end{abstract}

\pacs{}

\maketitle


\section{Work extraction through coupling to a work storage}
{The work extraction framework we use has been studied thoroughly in~\cite{AlhambraFluctuatingWorkQuantum2016}. 
We will summarize here some details of the underlying concept which are necessary for our scenario.}

{The system $\Sys$ and a work storage $\W$ with local Hamiltonians $H_\Sys$ and $H_\W$ are coupled with the aim to transfer work between the two. The work storage is in general assumed to be a continuous degree of freedom with
   $H_\W = \int x \kb{x}{x}\, dx$, with an orthonormal basis $\{\ket{x}, \forall{x} \in \mathds{R}\}$.}   

{The coupling is established by a joint unitary operation $\U_{\Sys\W}$. In order to avoid unaccounted energy contributions due to the coupling, the unitary has to be energy conserving, meaning it commutes with the global Hamiltonian
\begin{align}
	[\U_{\Sys\W}, H_\Sys\otimes \id + \id \otimes H_\W] = 0.
\end{align}
Since the Hamiltonians are purely local the change of energy in $\W$ due to the coupling $\U_{\Sys\W}$ has to be exactly compensated by the change of energy in $\Sys$
\begin{align}
	\Delta W_\W = -\Delta E_\Sys.
\end{align}
Furthermore, the whole process should be invariant under translations of the work storage system, that is $[\U_{\Sys\W}, \Delta_\W] = 0$, where $\Delta_\W$ is defined by $[H_\W, \Delta_\W] = i$. This condition is
necessary in order to avoid that the work storage can be used to lower the entropy of the system which would allow for an energy transfer in form of heat instead of work~\cite{AlhambraFluctuatingWorkQuantum2016,MasanesGeneralDerivationQuantification2017}.}

As shown in the Appendix of~\cite{AlhambraFluctuatingWorkQuantum2016} it is always possible to implement a global unitary $\U_{\Sys\W}$ which satisfies the conditions above and which acts \textit{locally} on $\Sys$ as a unitary $\U_\Sys$:
\begin{align}
	\Tr_\W\{ \U_{\Sys\W} (\rho_\Sys \otimes \rho_\W) \U_{\Sys\W}^\dagger\} =
	\U_\Sys \rho_\Sys \U_\Sys^\dagger. 
\end{align}

 {Bob achieves the largest work extraction for a given pure state $\ket{\psi_\Sys}$ if he transforms it to the ground state $\ket{0}_\Sys$ of his local Hamiltonian (he could not lower the energy of his system any further).
If he implements the unitary which maps $\U_\Sys \ket{\psi_\Sys} = \ket{0}_\Sys$ in the manner described above, he can be sure that the energy growth in the work storage is maximized and given by $\Delta W_\W = \bra{\psi_\Sys} H_\Sys \ket{\psi_\Sys} - \bra{0} H_\Sys \ket{0}$.}

 {It might be counter intuitive that the global unitary reduces to a local one although $\U_{\Sys\W} \neq \U_{\Sys} \otimes \U_\W$ (if $\U_{\Sys\W} = \U_{\Sys} \otimes \U_\W$ the unitary could in general not be energy conserving for every input state). As shown in~\cite{AlhambraFluctuatingWorkQuantum2016} this behavior can in general only be obtained if the work storage $\W$ is initialized in an eigenstate of the translation operator $\Delta_\W$. Thus, the state $\rho_\W = \kb{\psi_\W}{\psi_\W}$ is pure, and $\langle{x}|\psi_\W \rangle \neq 0, \forall x $.}

 {Even though this general approach is mathematically elegant, it is questionable how physical such an initial state of the work storage is. In particular the measurement of the energy change in the work storage (that's the quantity Bob is interested in) could be hard to implement because the measured energies in the work storage must fluctuate strongly as shown in~\cite{tajimaUncertaintyRelationsImplementation2018} and therefore the averaging would require a large amount of measured data. We therefore construct for the qubit example an explicit collision model which is able to approximate the work extraction to an arbitrary precision by discrete two-qubit interactions. Such an approach could in principle be implemented as a quantum circuit with current quantum technologies.}

\section{Collision model approach to work extraction}

 {The work extraction is implemented by performing a local unitary $\U_\Sys$ on the system. The change of the inner energy is then equal to the work
  performed (or absorbed) by the system. However, this energy change has
  to be compensated by an energy supply or storage. 
  Instead of a single work storage quantum system with a single continuous
  degree of freedom, as shown in the section above, we
  formulate the work extraction using a qubit collision model approach
  which reproduces the local unitary on the system and preserves the
  energy on
  average~\cite{LorenzoCompositequantumcollision2017,FilippovDivisibilityquantumdynamical2017,CampbellSystemenvironmentcorrelationsMarkovian2018}.
  Thus, the mean value of extracted work is equal to
  the change of the inner energy in the work medium of the engine as required. Our
  approach can be approximated to arbitrary precision by a finite number of
  two-qubit interactions.}  

The local unitaries $\U_\Sys$ implemented in the work storage cells of
the main text are rotations around the $y$-axis: 
\begin{align}
  \label{eq:usys}
  \U_\Sys = e^{-i \frac{\phi}{2} \sigma_y}.
\end{align}
In a first step we construct this unitary from a collision model. The
system interacts with a sequence of ancilla qubits. The interaction
between the system and a single ancilla is given by
\begin{align}
  \U_{\Sys\Anc} =e^{-i \Delta \phi (\sigma_{+}\otimes \sigma_{-} +
  \sigma_{-} \otimes \sigma_{+})}.
\end{align}
Each ancilla is initialized in the  $y$-eigenstate $\ket{+_y}$ =
$\frac{1}{2}(\ket{0} + i \ket{1})$.
The map for the state of the
system $\Sys$ after a single collision can be written as:
\begin{align}
  \rho_\Sys' &= \Tr_\Anc\{\U_{\Sys\Anc} (\rho_\Sys \otimes
               \rho_\Anc)\U_{\Sys\Anc}^\dagger \} \nonumber\\
  &=\Tr_\Anc\{\U_{\Sys\Anc} (\rho_\Sys \otimes
  \kb{+_y}{+_y})\U_{\Sys\Anc}^\dagger \}
\end{align}
We assume each collision to be weak and expand the interaction
to first order in $\Delta \phi$, hence we get
\begin{align}
  \rho_\Sys' = \rho_\Sys - i \frac{\Delta\phi}{2} [\sigma_y, \rho_\Sys].
\end{align}
In the limit of infinitesimal steps $\Delta\phi \rightarrow d\phi$ we
therefore obtain the desired local unitary evolution $\U_\Sys$.\\

\paragraph{The ancillas as a stochastic work storage}
--- The state of a single ancilla after its collision with the system
reads to first order in $\Delta \phi$:
\begin{align}
  \rho_\Anc' &= \Tr_\Sys\{\U_{\Sys\Anc} (\rho_\Sys \otimes
               \rho_\Anc)\U_{\Sys\Anc}^\dagger \} \nonumber\\
             &=\rho_\Anc + \Delta \phi \Tr\{\sigma_x \rho_\Sys\}
               \sigma_z \nonumber\\
  &= \rho_\Anc + \Delta \phi \, a\, \sigma_z,
\end{align}
where $a$ is the $x$-component of the state $\rho_\Sys$.
Thus, the $z$-component of the ancilla after the interaction contains
information about the $x$-component of the system before the
collision.

The Hamiltonian of each ancilla is given by $H_\Anc =  \sigma_z$.
Accordingly, the change of the average energy of a single ancilla due to
its collision with the work medium is given by
\begin{align}
 \overline{\Delta E} = \Tr\{ H_\Anc (\rho_\Anc'-\rho_\Anc) \} = \Delta
  \phi \, a.
\end{align}
As we have derived above, the system state evolves under the unitary
$\U_\Sys$ during the collisions. Thus, in a continuous limit ($\Delta
\phi \rightarrow d\phi$) the component $a$ becomes a function of
$\phi$.
Integrating over the average energy for all the ancillas then yields:
\begin{align}
  \overline{W} = \int a(\phi)\, d\phi.
\end{align}
The $\phi$-dependence of $a$ is given by $a(\phi) = x_\xi \cos(\phi) + z_\xi
\sin(\phi)$, where $x_\xi$ and $z_\xi$ are the initial $x$- and
$z$-component of the system state.
So, we have:
\begin{align}
  \overline{W} = \int_0^\phi a(\phi')\, d\phi' = z_\xi - z_\xi \cos(\phi) + x_\xi \sin(\phi).
\end{align}
This is exactly the change of the inner energy $\Delta E_\Sys$ under a
$\sigma_y$-rotation about an angle $\phi$.
Therefore, the constructed collision model is energy
conserving. To use the work, Bob can measure the ancillas which leads
to a stochastic amount of work which depends on how many ancillas are
measured to be in the $\ket{1}$ state. However, on average over many
work medium qubits, he can extract $\overline{W}$.

\section{Work extraction bound for a LHS ensemble - infinite temperature}
 {Before we derive the steering inequality for a finite temperature we consider the infinite temperature case.
 In this limit} ($\eta \rightarrow 0$) the local Gibbs
state is given by the fully mixed state $\rho_\Sys^{\textrm{Gibbs}} = \frac{\mathds{1}}{2}$.
We will calculate the average work which Bob can expect to extract
with his red cells from a  {LHS} ensemble
 {$\mathcal{F} = \{\rho_\xi; p_\xi\}$, where $p_\xi \geq 0$ and $\sum_\xi p_\xi =1$}. Bob assumes
that Alice might know the current state $\rho_{ {\xi}}$ of his system because
of classical correlations with the environment. If Alice indeed knows
the current $\rho_\xi$, the best she can do is to announce $\U^1_z$
whenever $z_\xi = \Tr\{\sigma_z \,\rho_\xi\} > 0$.  {These are the states which
give a positive work output}. For
$z_\xi = \Tr\{\sigma_z \,\rho_\xi\} \leq 0$ Alice announces $\U_z^0 = \id$ and
Bob will leave his system unchanged.

For suitably many runs the  work output averaged over the ensemble
$\{\rho_\xi; p_\xi\}$ reads
\begin{align}
    \label{eq:13}
    \overline{W}_z &=  \sum_{\xi,z_\xi>0} p_\xi  \bigg( \Tr\{H_\Sys \rho_\xi\}
                     -\Tr\{H_\Sys
  \U^1_z\rho_\xi{\U_z^1}^\dagger\}\bigg)  \nonumber\\
  &=\sum_{\xi,z_\xi>0} p_\xi \bigg(\frac{1+z_\xi}{2} -
    \frac{1-z_\xi}{2}\bigg) \nonumber\\
                  &=\sum_{\xi,z_\xi>0} p_\xi z_\xi \nonumber\\
                  &=\frac{1}{2}\sum_{\xi,z_\xi>0} p_\xi z_\xi  +
                    \frac{1}{2}\sum_{\xi,z_\xi\leq0} p_\xi (-z_\xi) \nonumber\\
    &=\sum_\xi p_\xi \frac{|z_\xi|}{2},
\end{align}
where the fourth line comes from
\begin{align}
 \sum_\xi p_\xi z_\xi= \sum_{\xi,z_\xi>0}  p_\xi z_\xi +\sum_{\xi,z_\xi\leq 0} p_\xi  z_\xi = 0 = \Tr\{\sigma_z \rho_\Sys\}.
\end{align}
We have to note that Bob does not know the ensemble $\{\rho_\xi;p_\xi\}$.

Bob assumes  that Alice's knowledge about his current state does not
depend on whether he has placed the qubit in a red or a blue cell. Therefore, he takes
the ensemble $\{\rho_\xi; p_\xi\}$ to be the same also for the work
extraction in the blue cells. Of course, Alice again would announce the
unitary which leads to  larger work extraction. Thus, she tells him to
perform $\U_x^+$ whenever $x_\xi = \Tr\{\sigma_x\,\rho_\xi\} > 0$ and
$\U_x^{-}$ otherwise.
For the work extraction in the blue cells we then find:
\begin{align}
    \label{eq:14}
  \overline{W}_x  &=\sum_{\xi,x_\xi>0} p_\xi  \bigg( \Tr\{H_\Sys \rho_\xi\}
                    -\Tr\{H_\Sys \U^+_x\rho_\xi{\U_x^{+}}^\dagger\}\bigg)\nonumber\\
                  &+\sum_{\xi,x_\xi\leq 0} p_\xi  \bigg( \Tr\{H_\Sys \rho_\xi\} -\Tr\{H_\Sys
                    \U^-_x\rho_\xi{\U_x^{-}}^\dagger\}\bigg)\nonumber\\
                  &=\sum_\xi p_\xi \frac{1}{2}(|x_\xi| - z_\xi)\nonumber\\
                  &=\sum_\xi p_\xi \frac{|x_\xi|}{2},  \end{align}
  where the last equality comes from $\sum_\xi p_\xi z_\xi = 0$.

  \section{Finite temperature case}
  
  In order to respect the $\langle \sigma_z \rangle$ expectation value
  of the Gibbs state we have:
  \begin{align}
 \eta &= \Tr\{\sigma_z\, \rho_\Sys^{\textrm{Gibbs}}\} = \sum_\xi p_\xi z_\xi
        \nonumber\\
      &= \sum_{\xi,z_\xi>0} p_\xi z_\xi + \sum_{\xi,z_\xi\leq 0} p_\xi z_\xi. 
  \end{align}
Thus, for the $z_\xi>0$ we get: 
  \begin{align}
  \label{eq:19}
 \frac{1}{2}&\sum_{\xi,z_\xi>0} p_\xi z_\xi = \frac{1}{2}\eta -
  \frac{1}{2}\sum_{\xi,z_\xi\leq 0} p_\xi z_\xi\\
  \phantom{\Rightarrow \frac{1}{2}}&\phantom{\sum_{\xi,z_\xi>0} p_\xi z_\xi}=\frac{1}{2}\eta +
  \frac{1}{2}\sum_{\xi,z_\xi\leq 0} p_\xi |z_\xi|
\end{align}
and we find for the red cells:
\begin{align}
  \label{eq:20}
  \overline{W}_z &= \sum_{\xi,z_\xi>0} p_\xi z_\xi \nonumber\\
  &=\frac{1}{2}\sum_{\xi,z_\xi>0} p_\xi z_\xi + \frac{1}{2}\sum_{\xi,z_\xi\leq 0}
    p_\xi |z_\xi| +\frac{\eta}{2}\nonumber \\
                 &=\frac{1}{2}\sum_\xi p_\xi (|z_\xi| + \eta)=
                   \frac{1}{2}\bigg(\eta + \sum_\xi p_\xi |z_\xi|\bigg).
\end{align}
The unitaries for the blue cells are given by:
\begin{align}
  \U_x^+=\begin{pmatrix}
 -\frac{\sqrt{1-\eta }}{\sqrt{2}} & \frac{\sqrt{\eta +1}}{\sqrt{2}} \\
 \frac{\sqrt{\eta +1}}{\sqrt{2}} & \frac{\sqrt{1-\eta }}{\sqrt{2}} \\
\end{pmatrix},
&& \U_x^- = \begin{pmatrix}
 \frac{\sqrt{1-\eta }}{\sqrt{2}} & \frac{\sqrt{\eta +1}}{\sqrt{2}} \\
 -\frac{\sqrt{\eta +1}}{\sqrt{2}} & \frac{\sqrt{1-\eta }}{\sqrt{2}} \\
\end{pmatrix}.
\end{align}
Then, the expected work output is
\begin{align}
  \label{eq:22}
  \overline{W}_x &= \sum_{\xi,x_\xi>0} p_\xi \bigg(\Tr\{H_\Sys \rho_\xi\} - \Tr\{H_\Sys
                  \U_x^+\rho_\xi{\U_x^+}^\dagger\}\bigg)\nonumber\\
  &+\sum_{\xi,x_\xi\leq 0} p_\xi \bigg(\Tr\{H_\Sys \rho_\xi\} - \Tr\{H_\Sys
                  \U_x^-\rho_\xi{\U_x^-}^\dagger\}\bigg)\nonumber\\
  &= \frac{1}{2} \sum_\xi p_\xi (z_\xi(1+\eta) +
    |x_\xi|\sqrt{1-\eta^2})\nonumber\\
  &= \frac{1}{2}\bigg(\eta + \eta^2 + \sum_\xi p_\xi  |x_\xi|\sqrt{1-\eta^2} \bigg).
\end{align}
To obtain the steering inequality in the finite temperature case we
calculate the expected classical work for the ratio $c =
\frac{1}{\sqrt{1-\eta^2}}$ of red and blue cells:
\begin{align}
  \label{eq:1}
  \overline{W}_{\textrm{cl}} &= \frac{\overline{W}_z + c \overline{W}_x}{1+c} \nonumber\\
  & = {\frac{1}{2} \frac{ \eta + c (\eta + \eta^2) +
    \sum_\xi p_\xi(|z_\xi| + |x_\xi|)}{1+c}} \nonumber\\
  & \leq {\frac{1}{2} \frac{ \eta + c (\eta + \eta^2) +
    \sqrt{2}}{1+c}} \nonumber\\
  &=\frac{\eta  \left(\sqrt{1-\eta ^2}+\eta +1\right)+\sqrt{2-2 \eta ^2}}{2 \left(\sqrt{1-\eta ^2}+1\right)}.
\end{align}

\section{Connection to Additive Convex Criteria}
 {Our steering inequality is an adaption of the well studied additive convex criteria \cite{CavalcantiExperimentalcriteriasteering2009}.
Each work extraction unitary defines an observable whose expectation values for a certain state $\rho_\xi$ are given by:
\begin{align}
\label{eq:observables}
    \langle W_z^1 \rangle &= z_\xi \nonumber \\
    \langle W_z^0 \rangle &= 0 \nonumber \\ 
    \langle W_x^+ \rangle &= \frac{1}{2} (z_\xi(1+\eta) + x_\xi \sqrt{1-\eta^2}) \nonumber \\
    \langle W_x^- \rangle &= \frac{1}{2} (z_\xi(1+\eta) - x_\xi \sqrt{1-\eta^2})
\end{align}
In contrast to the standard derivations of convex criteria we want to include the fact that the ensemble of states has to
respect the Gibbs state. Accordingly, instead of bounding a convex function of the observables~(\ref{eq:observables}) for each single $\rho_\xi$ we
directly look at the average over the whole ensemble under the assumption that Alice wants to maximize the work output (that's the best she can do):
\begin{align}
    \overline{W}_\textrm{cl} =& \frac{1}{1+c}\bigg( \sum_{\xi,z_\xi>0} p_\xi \langle W_z^1 \rangle + \sum_{\xi,z_\xi\leq 0} p_\xi \langle W_z^0 \rangle \nonumber \\
    &+ \frac{c}{2} (\sum_{\xi,x_\xi >0} p_\xi \langle W_x^+ \rangle + \sum_{\xi,x_\xi \leq 0} p_\xi \langle W_x^- \rangle)\bigg)
\end{align}
For $c=\frac{1}{\sqrt{1-\eta^2}}$ this value $\overline{W}_\textrm{cl}$ can be bounded as shown in the previous section.}

\section{Extractable Work under imperfect correlations }
We consider correlations between system and environment which are
described by the state:
\begin{align}
  \rho_{\Sys\Env} = q \rho_{\textrm{qu}} + (1-q) \rho_{\textrm{cl}}, 
\end{align}
as described in the main text. For the red cells Alice can help Bob to
still extract on average $\overline{W}_z = \frac{1+\eta}{2}$. For
the blue cells we now have $\overline{W}_x = \frac{1}{2}(q + \eta +\eta^2
- q \eta^2)$. Weighted with $c$, therefore, the average
amount of work will be:
\begin{align}
  \overline{W}_{\textrm{qu}} = \frac{(\eta +1) \left(\sqrt{1-\eta ^2}+\eta -\eta  q+q\right)}{2 \left(\sqrt{1-\eta ^2}+1\right)}.
\end{align}

\section{Quantum Szilárd engine simulation}
We will now construct an exemplary model, how the quantum correlations
Alice exploits can emerge from a thermalization process.
A suitable framework known from open quantum system theory is based on collision models~\cite{LorenzoCompositequantumcollision2017,FilippovDivisibilityquantumdynamical2017,CampbellSystemenvironmentcorrelationsMarkovian2018}. The dynamics -- the relaxation to a thermal state of Bob's
system -- is given by a sequence of short interactions between the
system and subenvironments. Initially each subenvironment is in the two-qubit state
\begin{align}
  \label{eq:17}
  \ket{\Psi}_{\Env\Env'} = \sqrt{\frac{1+\eta}{2}}\ket{11} + \sqrt{\frac{1-\eta}{2}}\ket{00}. 
\end{align}
The part $\Env$ interacts with the system by a joint unitary $ \Q_{\Sys\Env} = \exp[{-i \sqrt{\gamma \Delta t}\, \operatorname{SWAP}}]$,
where $\gamma$ is a positive coupling constant, $\Delta t$ is a short time
interval and $\operatorname{SWAP}$ is the two-qubit swap-gate.
The reduced dynamics for Bob's system is given by the one-collision
map
\begin{align}
  \label{eq:26}
  \rho_\Sys' = \Tr_{\Env\Env'}\{\Q(\rho_\Sys \otimes \kb{\Psi}{\Psi}_{\Env\Env'})\Q^\dagger\}.
\end{align}
A time-continuous limit may be obtained by expanding $\Q$ to first
order in $\Delta t$ and taking the limit $\Delta t \rightarrow dt$~\cite{CiccarelloCollisionmodelsquantum2017}.
The steady state in Bob's system is indeed the Gibbs
state~(\ref{eq:1}). 
Instead of tracing out the environment and discarding the
information, Alice couples the part $\Env'$ to another ancilla
qubit, called the control qubit $\Cont$~\footnote{She could also measure the subenvironments after the
interaction with Bob's system~\cite{BeyerCollisionmodelapproachsteering2018}. This would not change the reduced
dynamics on Bob's side.}. The interaction is taken to be the same as the one between
$\Sys$ and $\Env$, that is  $\Q_{\Env'\Cont} = \Q_{\Sys\Env}$.
The full unitary transformation in the collision model is then given by
\begin{align}
  \label{eq:27}
  \T_{\Sys\Env\Env'\Cont} =  \Q_{\Sys\Env}\otimes \id_{\Env'\Cont} +
  \id_{\Sys\Env} \otimes\Q_{\Env'\Cont},
\end{align}
and the one-collision map for the joint state $\rho_{\Sys\Cont}$ reads
\begin{align}
  \rho_{\Sys\Cont}' = \Tr_{\Env\Env'}\{\T(\rho_{\Sys\Cont} \otimes \kb{\Psi}{\Psi}_{\Env\Env'})\T^\dagger\}.
\end{align}
The steady state of this dynamics is the state
$\ket{\Psi}_{\Sys\Cont} = \ket{\Psi}_{\Sys\Env}$ which Alice can use to violate the steering inequality.
\begin{figure}
  \def\svgwidth{.5\columnwidth}
    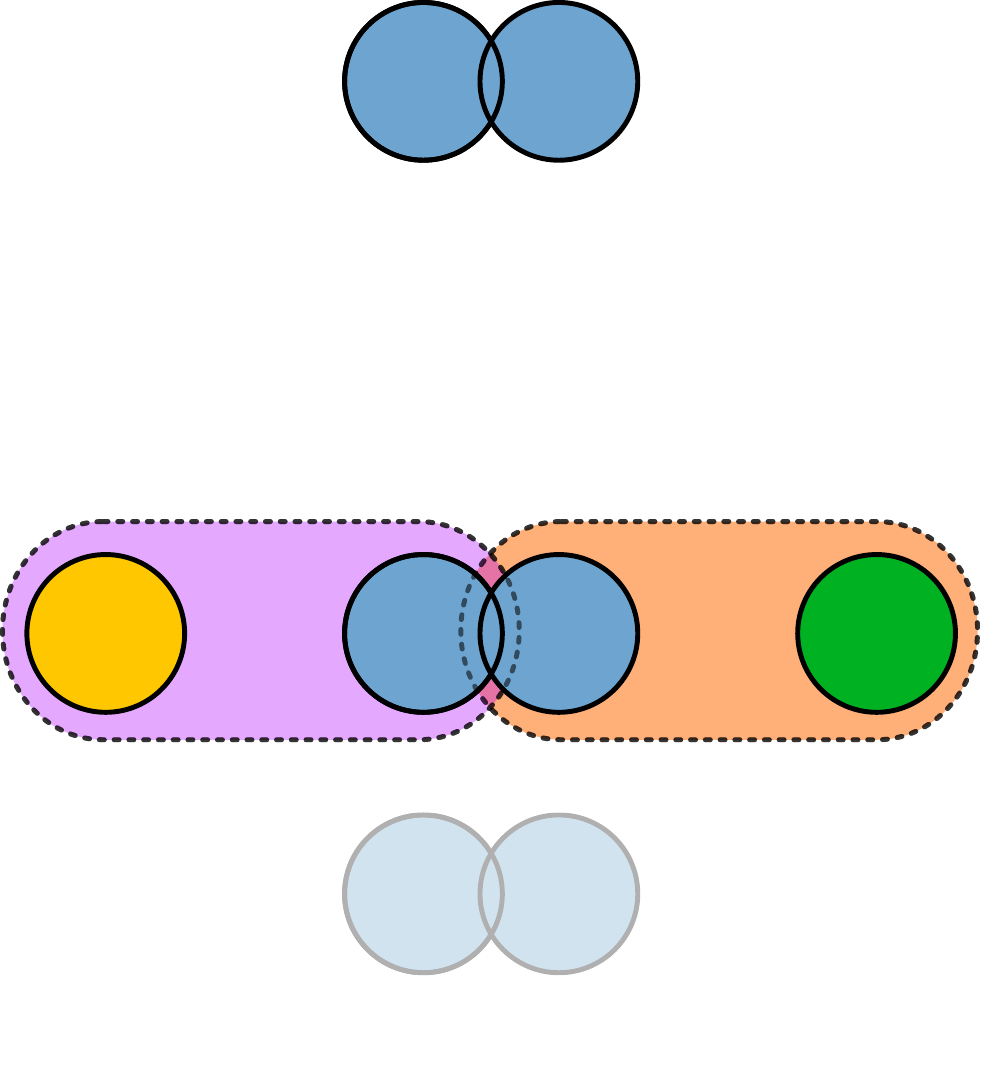
    \caption{Collision model which correlates Bob's system with
      Alice's control qubit. The
      subenvironments are two-qubit systems $\Env+\Env'$. $\Env$
      interacts with $\Sys$, $\Env'$ with $\Cont$. After the 
      thermalization the system is placed in either a red or a blue
      cell and Bob asks Alice which unitary he should use to
      extract work. For the two kinds of cells she measures the control qubit
      in different bases to answer this question.}\label{fig:cm}
  \end{figure}

\bibliography{bibliography}